\documentclass[aps, prb, showpacs,amsmath,amssymb, preprint, floatfix, superscriptaddress]{revtex4}
\usepackage{graphicx}
\usepackage{bm}
\begin{document}

\title{Pressure induced gap closing and metallization of  MoSe$_{2}$ and MoTe$_{2}$}

\author{Michaela Riflikov\'{a}}
\affiliation{Department of Experimental Physics, Comenius University, 
Mlynsk\'{a} Dolina F2, 842 48 Bratislava, Slovakia}
\affiliation{Institute of Electrical Engineering, Slovak Academy of Sciences, D\'{u}bravsk\'{a} cesta 9, 841 04 Bratislava, Slovakia}

\author{Roman Marto\v{n}\'{a}k}
\email{martonak@fmph.uniba.sk}
\affiliation{Department of Experimental Physics, Comenius University, 
Mlynsk\'{a} Dolina F2, 842 48 Bratislava, Slovakia}

\author{Erio Tosatti}
\affiliation{International School for Advanced Studies (SISSA) and CNR-IOM Democritos, Via Bonomea 265, I-34136 Trieste, Italy}
\affiliation{The Abdus Salam International Centre for Theoretical Physics
  (ICTP), Strada Costiera 11, I-34151 Trieste, Italy}


\date{\today}

\begin{abstract}
Layered molybdenum dichalchogenides are semiconductors whose gap is
controlled by delicate interlayer interactions. The gap tends to drop
together with the interlayer distance, suggesting collapse and
metallization under pressure.  We predict, based on first principles
calculations, that layered semiconductors 2H$_c$-MoSe$_2$ and
2H$_c$-MoTe$_2$ should undergo metallization at pressures between 28
and 40 GPa (MoSe$_2$) and 13 and 19 GPa (MoTe$_2$). Unlike MoS$_2$
where a 2H$_c$ $\to$ 2H$_a$ layer sliding transition is known to take
place, these two materials appear to preserve the original 2H$_c$
layered structure at least up to 100 GPa and to increasingly resist
lubric layer sliding under pressure. Similar to metallized MoS$_2$
they are predicted to exhibit a low density of states at the Fermi
level, and presumably very modest superconducting temperatures if
any. We also study the $\beta$-MoTe$_2$ structure, metastable with a
higher enthalpy than 2H$_c$-MoTe$_2$.  Despite its ready semimetallic
and (weakly) superconducting character already at zero pressure,
metallicity is not expected to increase dramatically with pressure.
\end{abstract}

\maketitle

\section{Introduction}

Transition Metal Dichalchogenides (TMDs) are well known and long
characterized compounds.\cite{Wilson_Yoffe, gmelin}.  They possess a
layered crystal structure consisting of MX$_2$ (M-transition metal,
X-chalcogen) XMX composite triatomic layers where the transition metal
monoatomic layer is sandwiched between two layers of chalcogen
atoms. The layers consisting of covalently bonded atoms are only
weakly coupled by partly van der Waals interactions, resulting in
highly anisotropic properties. By analogy with graphite, their
structure based on independently stable, relatively unreactive
triatomic layers which can be mutually sheared is probably related to
the functioning of some materials such as MoS$_2$ and MoSe$_2$ as
lubricants. It opens at the same time the way to a rich polytypism,
due to various possible relative stackings of the layers. Much initial
interest in bulk materials has been driven by their electronic
properties, including insulator, semiconductor, metal,
charge-density-wave (CDW) material, and superconductor -- properties
that can also be modified by intercalation.  More recently, focus
shifted to the exfoliated monolayers, similar to graphene.  Owing to
removal of interlayer interactions a monolayer has, unlike the bulk
material, a larger and direct band gap, features which in MoS$_2$ make
it of interest for optoelectronics
(Refs.\cite{PhysRevLett.105.136805}, \cite{NaturePhotonics},
\cite{Yue20121166}, \cite{tabatabaei163708}, \cite{LuPhysChem},
\cite{Shi_PhysRevB.87.155304}).

On the opposite front, it is possible to modify the properties of bulk
TMDs by external hydrostatic pressure which can in principle cause
structural as well as electronic phase transitions.  In
Ref.\cite{aksoy-MoSe2} 2Hc-MoSe$_2$ was compressed up to 35.9 GPa and
studied by X-ray diffraction but no structural transition was
reported.  At normal conditions MoS$_2$, MoSe$_2$ and MoTe$_2$ are
semiconductors with indirect energy gaps of about 1.29 eV, 1.1 eV and
1.0 eV, respectively.

The behaviour of MoS$_2$ under pressure is now well understood.  Its
resistivity decreases with pressure\cite{resistivity1} suggesting
possible metallization at high pressure. While that possibility is
confirmed by calculations~\cite{hromadova} it was also theoretically
found that the initial 2H$_c$ structure (hexagonal, space group
$P6_{3}/mmc$) at zero pressure should undergo near 20 GPa a pressure
induced structural transition to 2H$_a$ (the same space group
$P6_{3}/mmc$), the structure typical of e.g. NbS$_2$. That result
explained some previously mysterious X-ray diffraction
evidence\cite{aksoy} and Raman scattering data\cite{Livneh2010} and
was also recently confirmed experimentally
\cite{bandaruetal2014}. Both MoS$_2$ phases were predicted to
metallize at the same pressure region where the structural transition
takes place, 2H$_c$ at 25 GPa and 2H$_a$ at 20 GPa -- an electronic
result not yet investigated by experiments.

The behaviour of MoS$_2$ suggests the possibility that
pressure-induced gap closing similar to that of MoS$_2$ might occur in
the similar materials MoSe$_2$ and MoTe$_2$.  The experimental
electrical resistivity of MoSe$_2$ under pressure appears to be
controversial. In Ref.\cite{resistivity1} a sudden resistivity drop
was found at 4 GPa, with no interpretation provided.  In more recent
work Ref.\citep{resistivity2} resistivity was studied up to 8 GPa and
found to decrease smoothly upon compression with no sudden drop.

For the next member of this group, MoTe$_2$, also semiconducting in
its 2H$_c$ room temperature, zero pressure stable form
($\alpha$-MoTe$_2$), the pressure dependence of resistivity is not
known.  High temperature is known to induce a structural transition to
a new $\beta$-MoTe$_2$ phase\cite{Brown1966} (monoclinic, space group
$P2_{1}/m$), still a layered structure with additional modulation of
structure inside the layers, so that the Mo atoms now present a
distorted octahedral coordination rather than the trigonal prismatic
one of 2H$_c$ \cite{Dawson_Bullett}.  Interestingly the $\beta$-phase
is metallic at zero pressure.  The transition from $\alpha$- to
$\beta$-MoTe$_2$ occurs by raising temperature to
$900\,^{\circ}\mathrm{C}$ (Ref.\cite{Brown1966}) but the new structure
survives in a metastable state upon cooling down to room temperature
-- or even to cryogenic temperature where it reportedly shows
superconductivity~\cite{Wilson_Yoffe}. No high-pressure data appear
to be available for either $\alpha$- or $\beta$-MoTe$_2$, and we can
therefore only rely on theory concerning their structural and
electronic behaviour in that regime.

Here we present first principles calculations based on density
functional theory (DFT) demonstrating the effect of high pressure on
bulk transition metal dichalcogenides MoSe$_2$ and MoTe$_2$, focusing
both on the evolution of crystal structure and of electronic
properties. We first of all will describe in the next section the
technical details of DFT calculations. The following section will
present our predicted evolution of crystalline and electronic
structure, predicting the absence of structural transformations,
surprising in view of the initial analogy to MoS$_2$, and a
semiconductor-band overlap metal transition for both MoSe$_{2}$ and
MoTe$_{2}$ upon compression.  After a discussion of similarities and
differences with MoS$_2$, in particular of the similarly poor metallic
properties at high pressure, the last section will summarize our
findings and draw conclusions.

\begin{figure}[htpb]
  \includegraphics*[width=8.8cm]{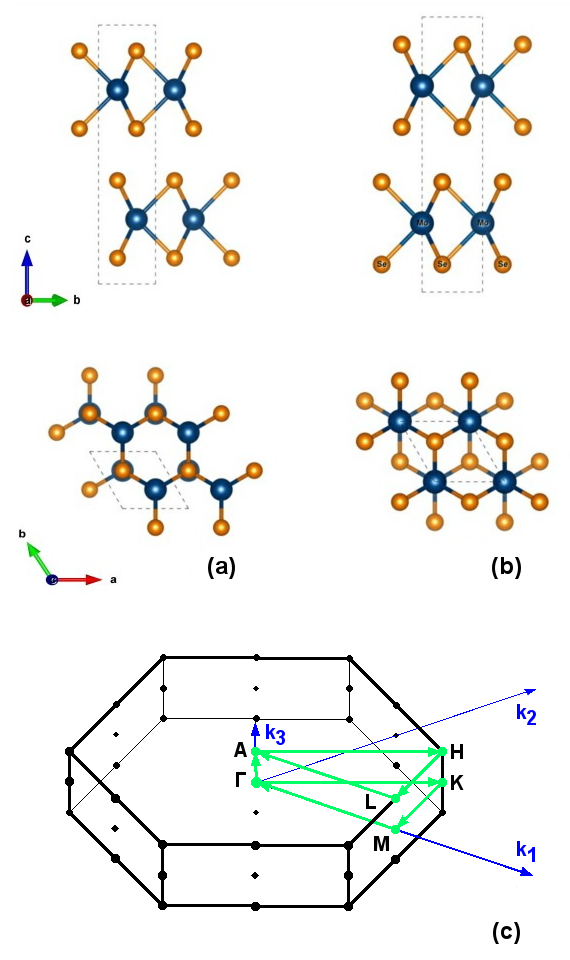}
  \caption{Structures of 2H$_c$-MoSe$_2$ (a) and 2H$_a$-MoSe$_2$ (b).
    I.~Brillouin zone for both structures\cite{xcrysden} (c).}
 \label{fig:structures}
\end{figure}

\section{Calculation method}

We follow well established understanding of, and specifically our own
fresh experience with,\cite{hromadova} layered TMDs. Straight density
functional total energy and structural calculations, quite delicate
and uncertain at zero pressure due to large effects caused by the
otherwise weak long-range interlayer dispersion van der Waals (vdW)
forces which are beyond simple DFT, become much more reliable and
predictive at high pressures, where vdW corrections become
unnecessary. The case of MoS$_2$ had been particularly instructive in
this respect, showing that whereas the calculated zero-pressure c-axis
interlayer spacing with the simple PBE exchange-correlation functional
(no vdW) was as expected larger than experiment, it improved
substantially at 5 GPa, turning extremely close to experiment at 10
GPa and upwards. Assuming, as is very reasonable, the same to hold for
MoSe$_2$ and MoTe$_2$, we used no vdW correction and restricted our
calculations to pressures of 10 GPa and higher, with no attempt to
explore the more delicate and less interesting low pressure region.
In order to increase the dependability of our predicted metallization
pressures, we also repeated the electronic structure calculations
(with PBE optimized structures, which are trustworthy) with the HSE06
functional which, contrary to PBE, overestimates band gaps and
therefore metallization pressures too.  The HSE06 calculations were
used to establish an upper bound to the semiconductor-metal transition
pressure.  We employed the Quantum ESPRESSO package\cite{QE} for
structural optimization and calculation of electronic properties. We
used scalar relativistic
pseudopotentials \footnote{Mo.pbe-spn-kjpaw\_psl.0.2.UPF,
  Se.pbe-n-kjpaw\_psl.0.2.UPF and Te.pbe-dn-kjpaw\_psl.0.2.2.UPF from
  http://www.quantum-espresso.org} and similarly to
Ref.\cite{hromadova} we employed a PBE exchange-correlation
functional\cite{pbe} with cutoff of 950 eV.  For 2H$_c$ structures we
used the six-atom unit cell and $17 \times 17 \times 5$ Monkhorst -
Pack\cite{mpgrid} (MP) \textit{k}-point sampling grid for relaxations
and $24 \times 24 \times 8$ MP grid for electronic DOS
calculations. For the $\beta$-MoTe$_2$ structure we used the unit cell
with 12 atoms and k-points grids $7 \times 15 \times 5$ and $12 \times
24 \times 6$ for relaxation and DOS calculations,
respectively. Spin-orbit coupling was not included in the calculation
of total energy and structural optimizations, to which it contributes
only in second order. We instead carried out test calculations to
check the impact of spin-orbit on metallization pressures but found it
to be also negligible.  In all results presented below spin-orbit
interaction is therefore omitted. Hybrid functional calculations
employing the HSE06 functional were performed with norm-conserving
pseudopotentials \footnote{Mo.pbe-mt\_fhi.UPF, Se.pbe-mt\_fhi.UPF and
  Te.pbe-mt\_fhi.UPF from http://www.quantum-espresso.org}. Zero
temperature and neglect of zero-point energy contributions were
assumed throughout.
 
A series of  PBE calculations were carried out at increasing pressures from
10 GPa upwards, with at each pressure a full structural relaxation
aimed at identifying the minimum enthalpy structure, its electronic
band structure, and their pressure evolution.

\section{Results}

\subsection{MoSe$_2$}

\begin{figure}[htpb]
  \includegraphics*[width=15cm]{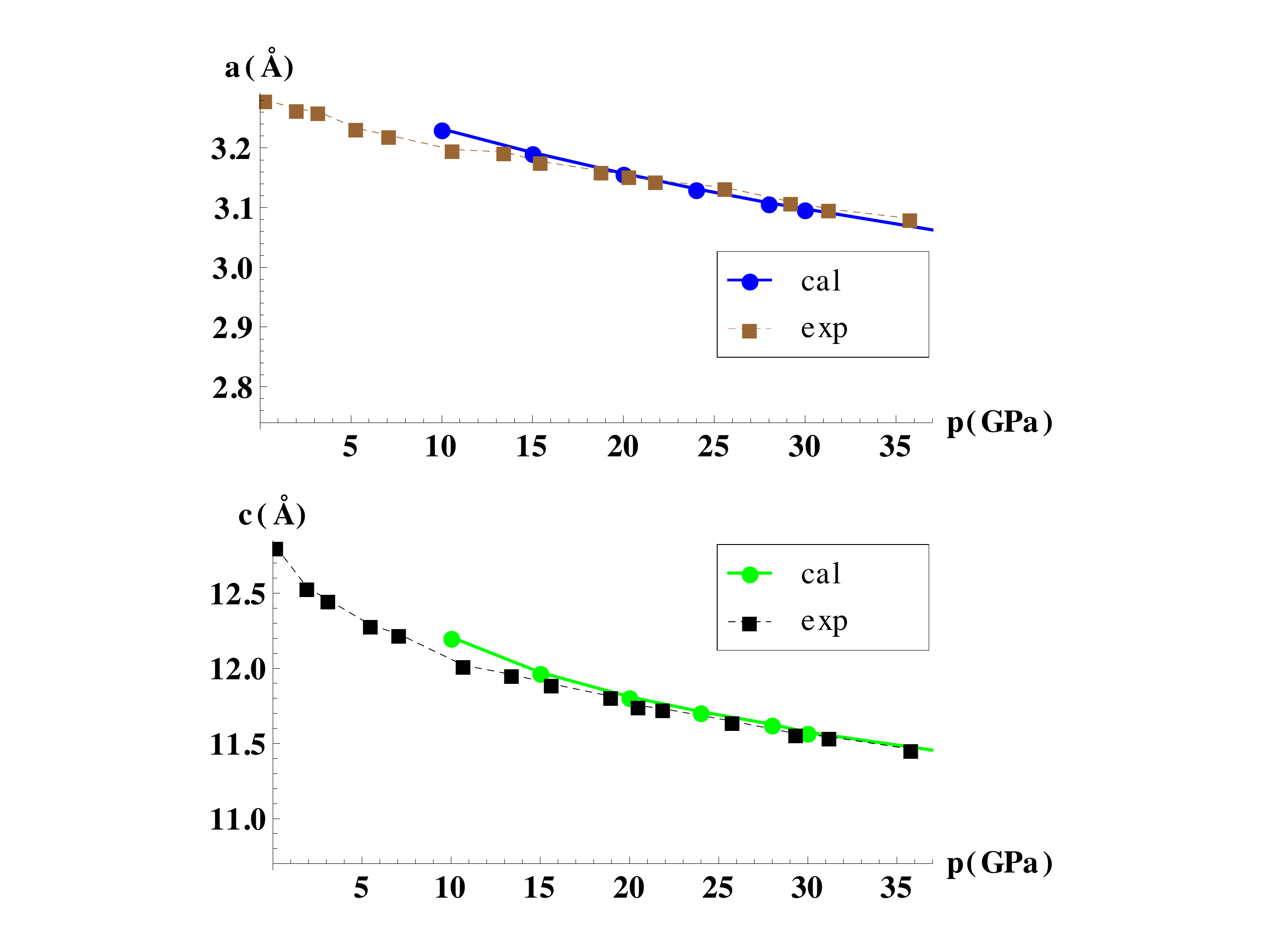}
  \caption{Pressure dependence of the calculated lattice parameters
    $a$ (upper panel) and $c$ (lower panel) of 2H$_c$-MoSe$_2$
    together with experimental data from Ref.\cite{aksoy-MoSe2}.}
 \label{fig:lattice_par}
\end{figure}

We performed a compression of the 2H$_c$-MoSe$_2$ unit cell up to 130
GPa and calculated the pressure dependence of the lattice parameters
$a$ (intra-layer), and $c$ (inter-layer)
(Fig.\ref{fig:lattice_par}). For comparison we show in the same figure
the experimental data extracted from X-ray diffraction patterns in
Ref.\cite{aksoy-MoSe2}.  As can be seen the agreement is excellent
especially at pressures beyond 15 GPa, which justifies {\it a
  posteriori} the use of PBE functional without vdW corrections, as
was also the case in MoS$_2$\cite{hromadova}.

\begin{figure}[htpb]
  \includegraphics*[width=8.8cm]{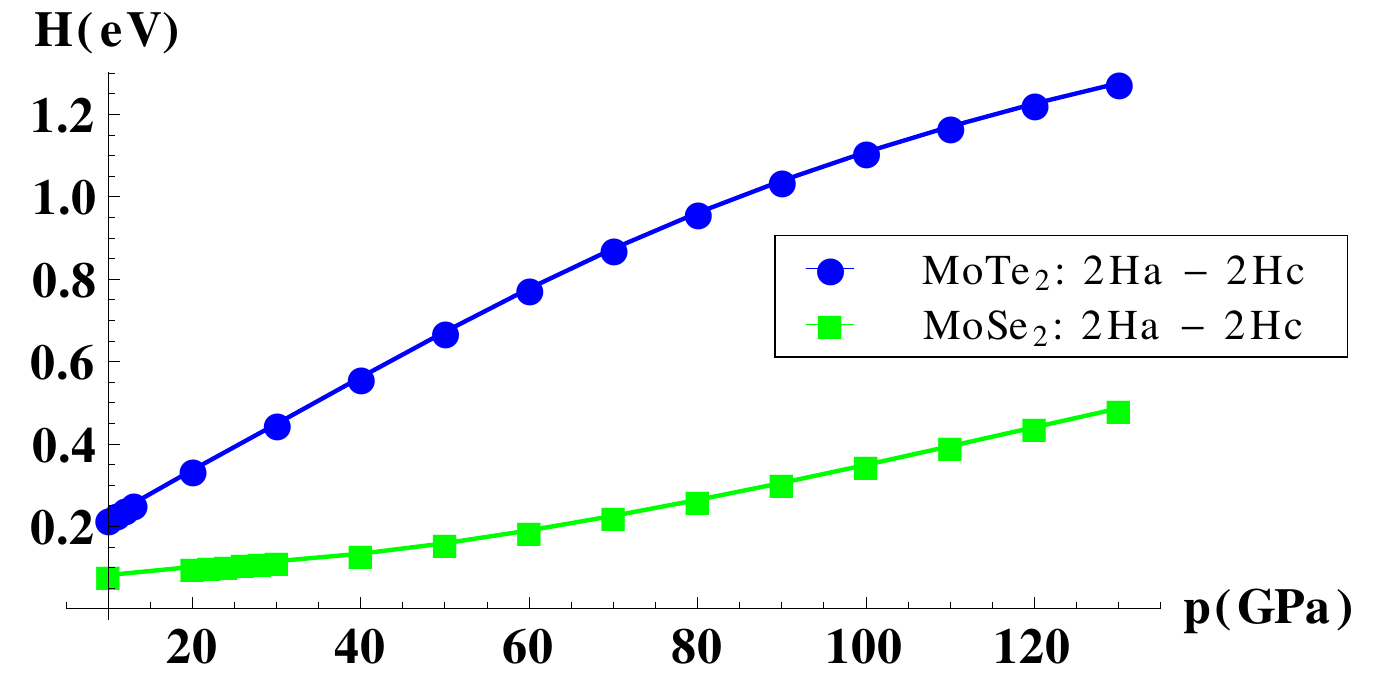}
  \caption{Pressure dependence of the enthalpy difference between the
    2H$_a$ and 2H$_c$ structures of MoSe$_2$ and MoTe$_2$,
    respectively. Note the increase with pressure, indicating
    increased stability of 2H$_c$ in both materials, unlike MoS$_2$
    where 2H$_a$ prevails at high pressure\cite{hromadova}. }
 \label{fig:enthalpies}
\end{figure}

Results also agree with experiment Ref.\cite{aksoy-MoSe2} in MoSe$_2$
indicating no structural changes and the stability of the 2H$_c$ zero
pressure structure of MoSe$_2$ at least up to 35.9 GPa. In order to
check whether any transition could take place at a higher pressure
than this, it would be necessary in the future to conduct some kind of
structural search. Limiting ourselves to explore the simple
possibility of a transition to the 2H$_a$ structure, we calculated the
enthalpies of both 2H$_a$ and 2H$_c$ phases of MoSe$_2$ up to 130
GPa. Fig.~\ref{fig:enthalpies} shows the enthalpy difference between
the two phases. Unlike the case of MoS$_2$ where the enthalpies cross
and the 2H$_a$ polytype became more stable around 20 GPa
\cite{hromadova}, here the enthalpy difference actually {\it
  increases} with pressure, thus reinforcing the stability of the
2H$_c$ structure. The slope of the enthalpy difference with increasing
pressure (Fig.\ref{fig:enthalpies} and Fig.2 in Ref.\cite{hromadova})
is equal to the volume difference between the respective phases and
the behaviour of MoSe$_2$ and MoTe$_2$ is just the opposite of
MoS$_2$: e.g. at p=20 GPa the volume of the unit cell of the 2H$_a$
phase compared to the 2H$_c$ one is larger in MoSe$_2$ and MoTe$_2$ by
0.2 \% and 1.4 \%, respectively, while in MoS$_2$ it is smaller by 1
\%. The layer-sliding structural transition observed in MoS$_2$
therefore is not expected to occur in MoSe$_2$ -- and, as we shall see
further down, neither is it in MoTe$_2$.

We can rationalize the reason for that difference of behaviour between
MoSe$_2$ or MoTe$_2$ and MoS$_2$ based on simple considerations of
interlayer Mo-Mo metallic bonding.  We first note that only in the
2H$_a$ structure the Mo atoms in nearest layers are vertically on top
of one another, whereas they are staggered and chemically far away in
2H$_c$. The 2H$_a$ structure can be energetically favored if it can
take advantage of d-electron propagation and metallicity along the
c-axis, such as is the case in NbS$_2$, NbSe$_2$, and high pressure
MoS$_2$. In MoSe$_2$ and MoTe$_2$, due to the larger radius of anions,
the interlayer Mo-Mo distances are larger, e.g., by about $0.25 \;\AA$
in MoSe$_2$ than in MoS$_2$. That makes interlayer d-electron
propagation energetically less important in MoSe$_2$ leaving
anion-anion repulsive forces in control of the enthalpy balance and
finally favoring 2H$_c$ over 2H$_a$.

\begin{figure}[htpb]
  \includegraphics*[width=15cm]{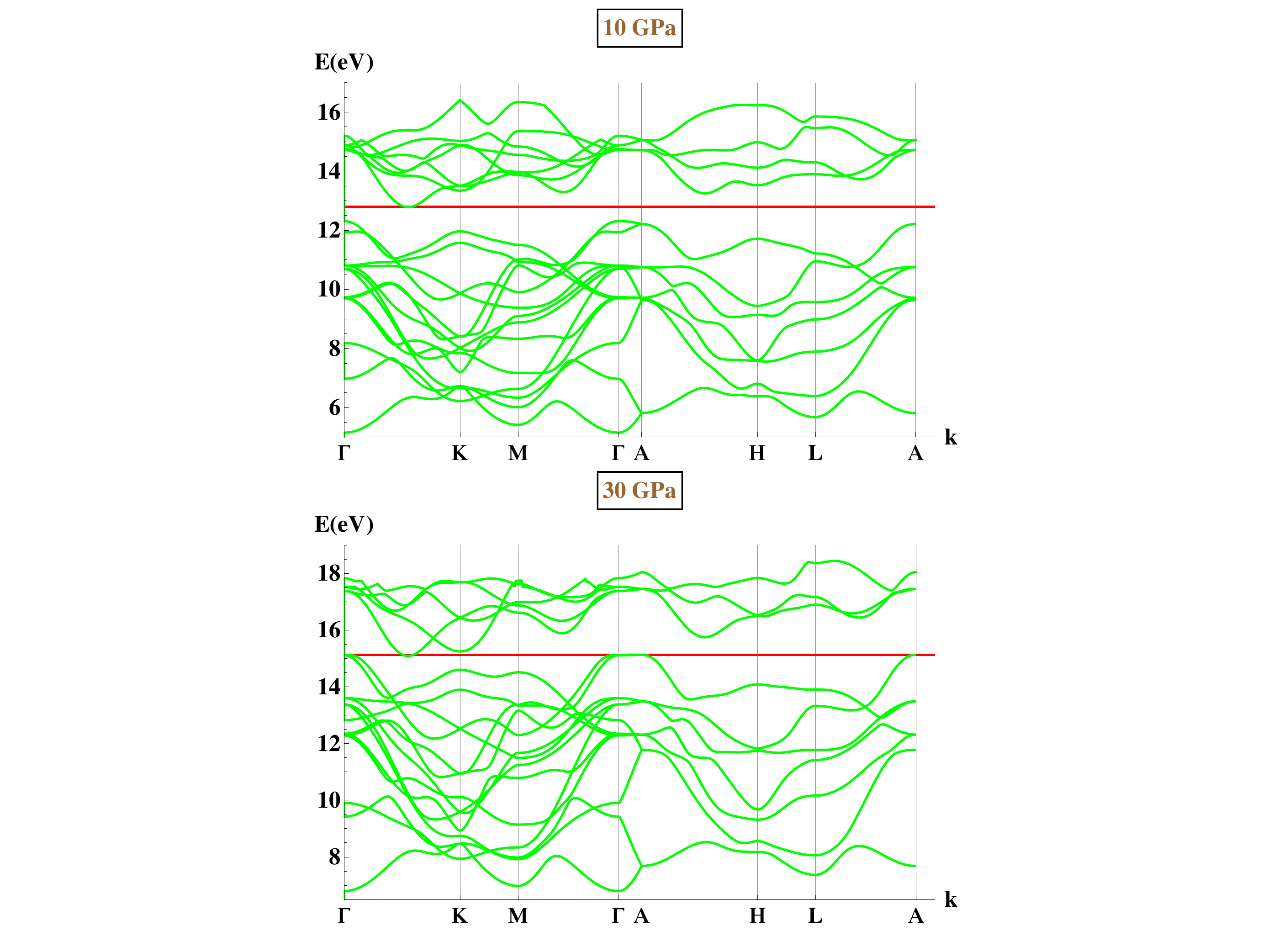}
  \caption{Band structure of 2H$_c$-MoSe$_2$ at $p=10$ GPa (upper
    panel) and $p=30$ GPa (lower panel).}
 \label{fig:bands_MoSe2}
\end{figure}

Having thus characterized the pressure evolution of the atomic
structure, we can examine that of electronic properties, in particular
the pressure induced closing of band gap and metallization.  The
calculated PBE band structure is shown in Fig.\ref{fig:bands_MoSe2}
for 2H$_c$-MoSe$_2$ at $p=10$ GPa and $p=30$ GPa. The gap decreases
with pressure at the rate of 0.026 eV/GPa. At $p=10$ GPa there still
is an indirect band gap of 0.47 eV with the valence band top at the
$\Gamma$ point and the conduction band bottom at some point $Q$ close
to the midpoint between $\Gamma$ and $K=\frac{1}{3} \mathbf{b}_{1} +
\frac{1}{3} \mathbf{b}_{2}$, where $\mathbf{b}_{1}, \mathbf{b}_{2}$
are reciprocal lattice vectors and $|\Gamma K| = \frac{4 \pi}{3 a}$.

\begin{figure}[htpb]
  \includegraphics*[width=8.8cm]{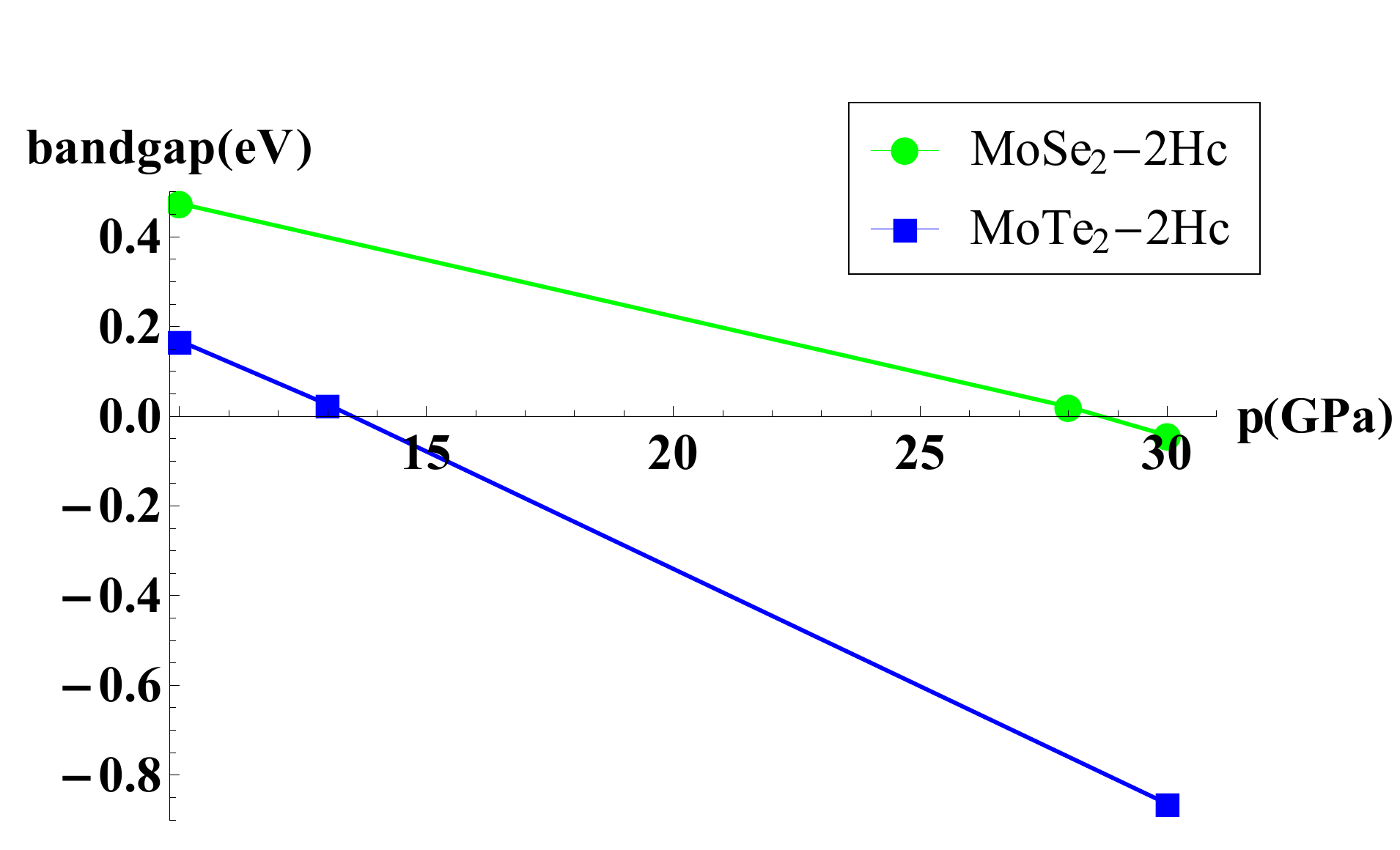}
  \caption{Pressure dependence of the band gap of 2H$_c$-MoSe$_2$ and 2H$_c$-MoTe$_2$.}
 \label{fig:band_gap_MoSe2}
\end{figure}

\begin{figure}[htpb]
 \includegraphics*[width=15cm]{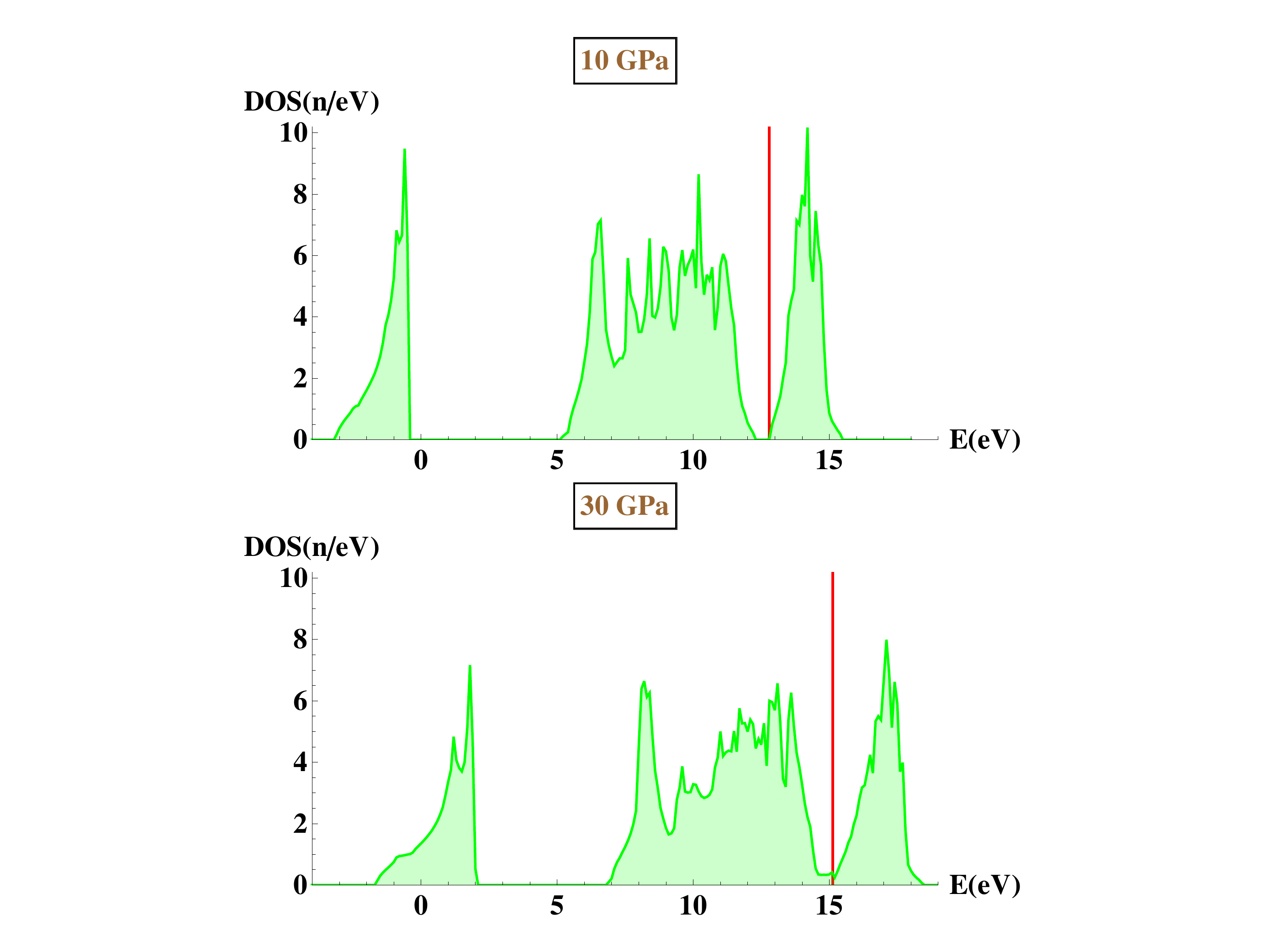}
  \caption{Density of states per unit cell of 2H$_c$-MoSe$_2$ at $p=10$ GPa and $p=30$ GPa.}
 \label{fig:dos_MoSe2}
\end{figure}

At $p=30$ GPa the band gap is already closed and the valence and
conduction bands exhibit a tiny overlap. Since the PBE approximation
certainly does not overestimate the band gap, our calculation suggests
that metallization of 2H$_c$-MoSe$_2$ at lower pressures than 30 GPa
is unlikely. This is compatible with the more recent resistivity data
of Ref.\cite{resistivity2} ( which appear to correct earlier results
\cite{resistivity1} which had suggested a transition at 4 GPa for
which there is no supporting evidence). In
Fig.\ref{fig:band_gap_MoSe2} we show our predicted pressure dependence
of the band gap which shows metallization by band overlap in MoSe$_2$
at $P_{met}$= 28 GPa.  Following band overlap, 2H$_c$-MoSe$_2$ turns
semimetallic with a low density of states at the Fermi level, as shown
by Fig.\ref{fig:dos_MoSe2} at $p=30$ GPa. To establish an upper bound
for the metallization pressure we then performed calculations using
the same structures, but the HSE06 hybrid functional\cite{hse06} in
place of PBE and found the gap closing at 40 GPa. Since this
approximation is known to overestimate the band gap, we conclude that
MoSe$_2$ metallizes at pressure between 28 and 40 GPa.

Since for an indirect band gap semiconductor the exciton binding
energy $E_B$ remains finite right up to $P_{met}$, there is in
principle the possibility upon gap closing to realize a so-called
excitonic insulator state.  That is a kind of charge-density-wave or
spin-density-wave state with wavevector $\mathbf{Q}$ theoretically
predicted long ago in Ref.\cite{PhysRev.158.462}. In a narrow range of
pressures immediately below $P_{met}$ the semiconducting gap becomes
small enough to be comparable with the exciton binding energy $E_B$,
expected here to be of order of 10 meV. Given a gap reduction rate of
26 meV/GPa, this means that the excitonic state could exist in a
narrow pressure range of about 4 kbar below $P_{met}$.  The DFT-PBE
electronic structure approximation does not treat properly the
nonlocal exchange which is essential for the description of excitons,
and thus it does not describe excitonic insulators states. Therefore
we cannot make a quantitative prediction of the relevant portion of
the pressure phase diagram and we must limit ourselves to a
qualitative statement. The possible realization of this interesting
state in MoS$_2$ was proposed in Ref.\cite{hromadova}, but the
structural transformation occurring at a pressure close to
metallization presents a fatal complication in that system.  From this
point of view MoSe$_2$ (and as we shall see also MoTe$_2$), structurally
stable lattice in the metallization region, appears to be a more
suitable system to search for an excitonic insulator state.

\subsection{MoTe$_2$}

\begin{figure}[htpb]
\includegraphics*[width=15cm]{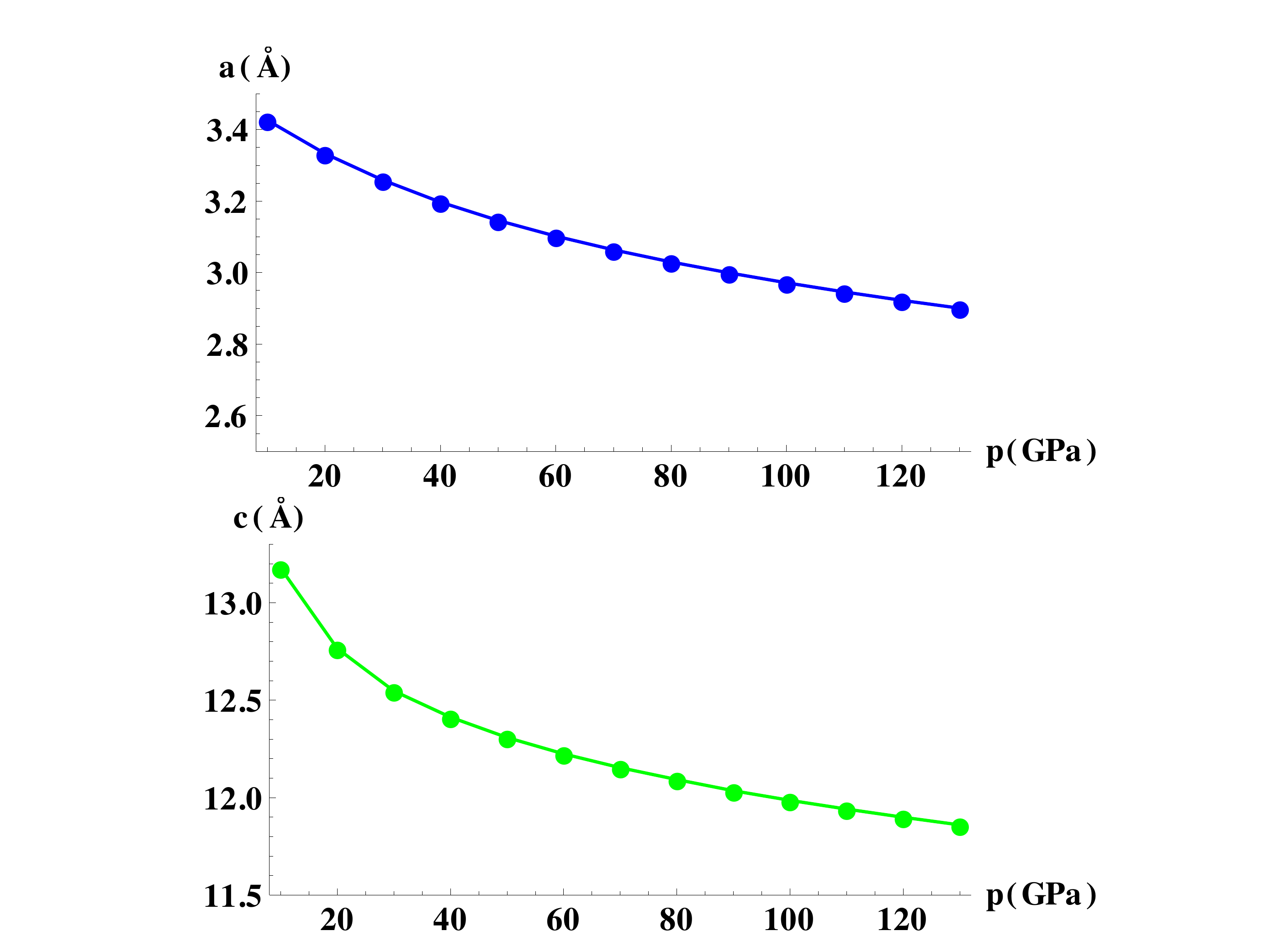}
  \caption{Pressure dependence of the calculated lattice parameters
    $a$ (upper panel) and $c$ (lower panel) of 2H$_c$-MoTe$_2$.}
 \label{fig:lattice_MoTe2}
\end{figure}

Compared with 2H$_c$-MoSe$_2$, there is much less experimental work
for 2H$_c$-MoTe$_2$ ($\alpha$-form), and we are not aware of either
structure or resistivity data under pressure, and our results
represent a first theoretical exploration.  We carried out the same
calculation protocol as for for 2H$_c$-MoSe$_2$: total energy
calculation, structural relaxation, enthalpy calculation, band
structure and gap calculation.  The calculated structural data are
shown in Fig.\ref{fig:lattice_MoTe2}.  Here too the 2H$_c$ structure
remains stable under pressure, at least with respect to a
transformation to 2H$_a$. The enthalpy difference stabilizing 2H$_c$
over the 2H$_a$ structure shown in Fig.\ref{fig:enthalpies} is here
even stronger than in MoSe$_2$, in agreement with our earlier
explanation involving the volume difference between the phases and the
larger radius of Te anions relative to Se.

\begin{figure}[htpb]
\includegraphics*[width=15cm]{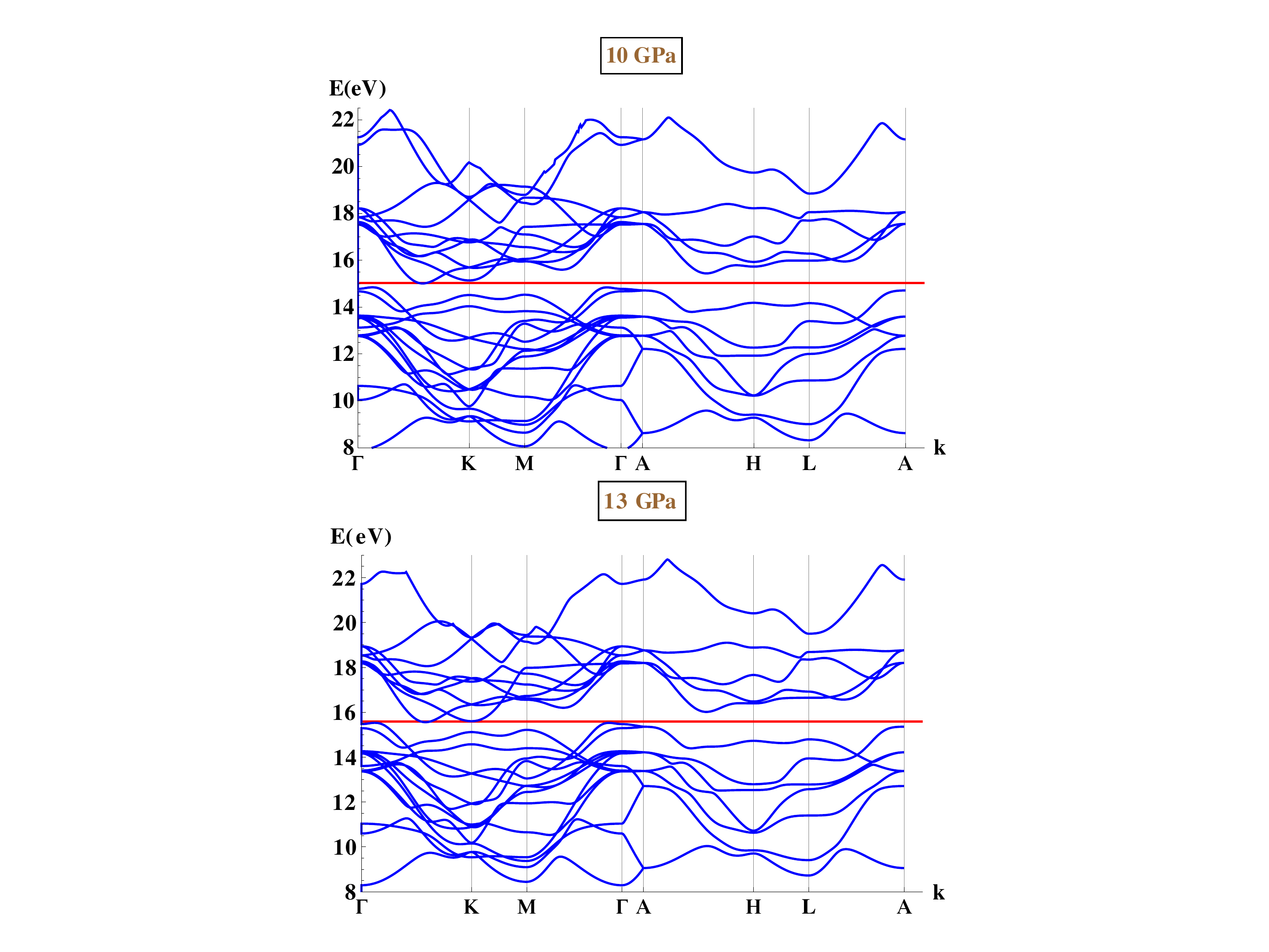}
  \caption{Band structure of 2H$_c$-MoTe$_2$ at $p=10$ GPa (upper panel) and $p=13$ GPa (lower panel).}
 \label{fig:bands_MoTe2}
\end{figure}

\begin{figure}[htpb]
  \includegraphics*[width=15cm]{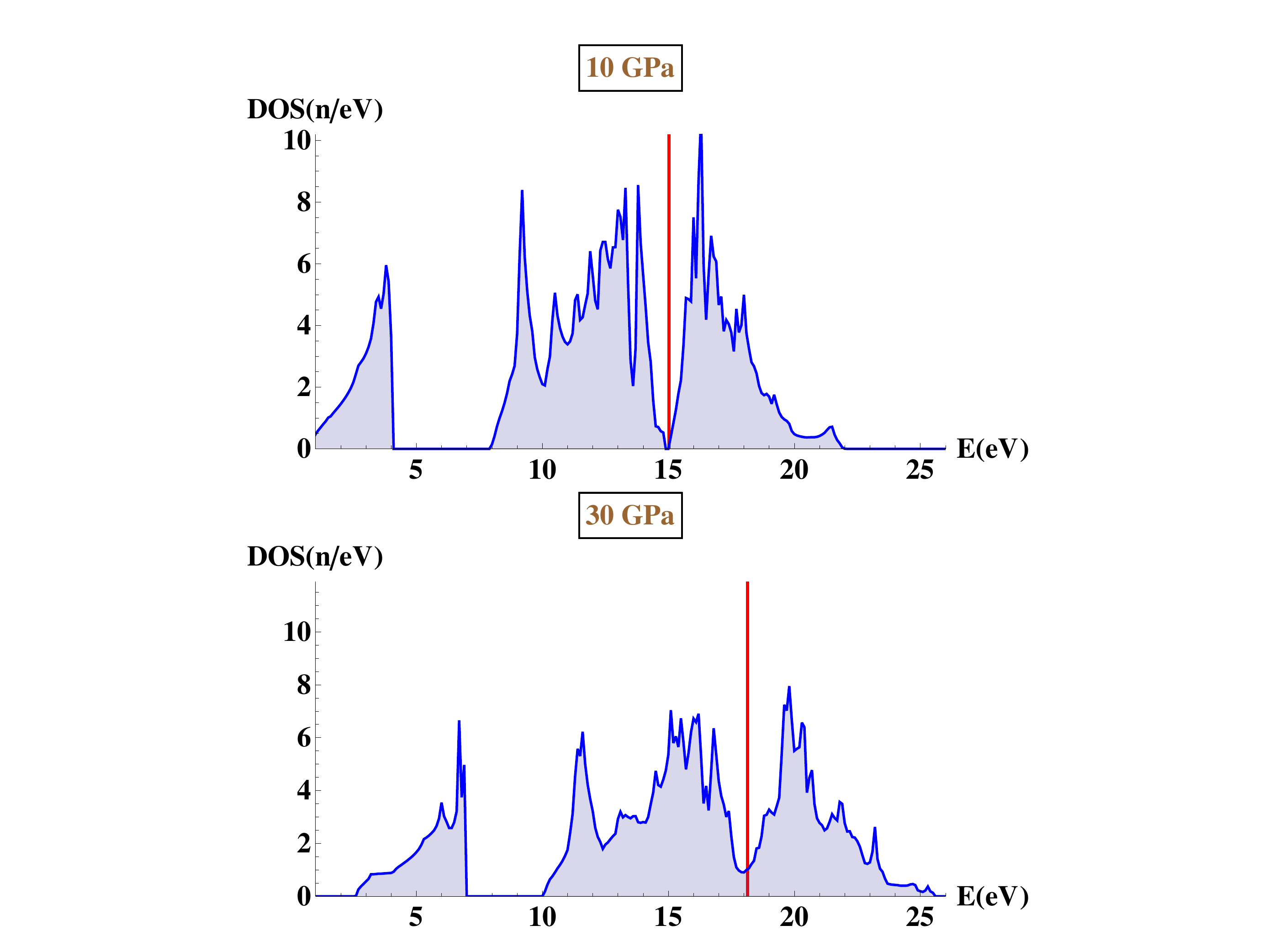}
  \caption{Density of states per unit cell of 2H$_c$-MoTe$_2$ at $p=10$ GPa and $p=30$ GPa.}
 \label{fig:dos_MoTe2}
\end{figure}

Fig.\ref{fig:bands_MoTe2} shows the band structure of
2H$_c$-MoTe$_2$. Here we took special care to verify that spin-orbit
interaction has no major effect on the states in the vicinity of the
Fermi level.  At 10 GPa there is still a small but finite band gap. At
13 GPa band overlap has already taken place between the valence band
top, now slightly displaced from $\Gamma$, and the conduction band
bottom which has two nearly degenerate minima - one again close to
midpoint $Q$ between $\Gamma$ and $K$ points and another one at the
$K$ point. Thus even in 2H$_c$-MoTe$_2$ there could be a narrow
excitonic insulator phase just below the metallization pressure;
however the CDW or SDW condensate wavevector is less straightforward
to predict.  Fig.\ref{fig:dos_MoTe2} shows the electronic density of
states at 10 and 30 GPa and one can see that even at 30 GPa, more than
twice of the metallization pressure, the electronic DOS remains rather
low, indicating semi-metallicity. Here again we performed a hybrid
functional\cite{hse06} calculation and found gap closing at 19 GPa,
thus placing the metallization pressure of 2H$_c$-MoTe$_2$ between 13
and 19 GPa.

\begin{figure}[htpb] 
  \includegraphics*[width=8.8cm]{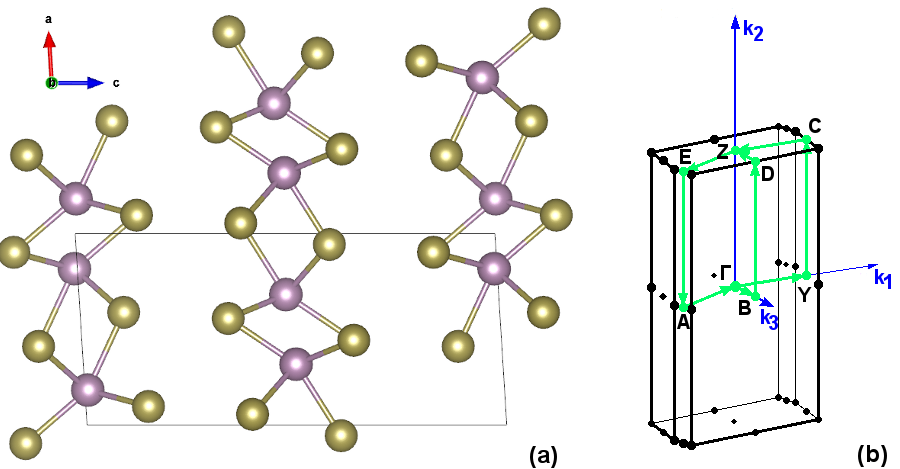}
  \caption{Structure of the $\beta$-form of MoTe$_2$ (a) and its
    I. Brillouin zone (b).}
 \label{fig:structure_beta_MoTe2}
\end{figure}

\begin{figure}[htpb] 
  \includegraphics*[width=15cm]{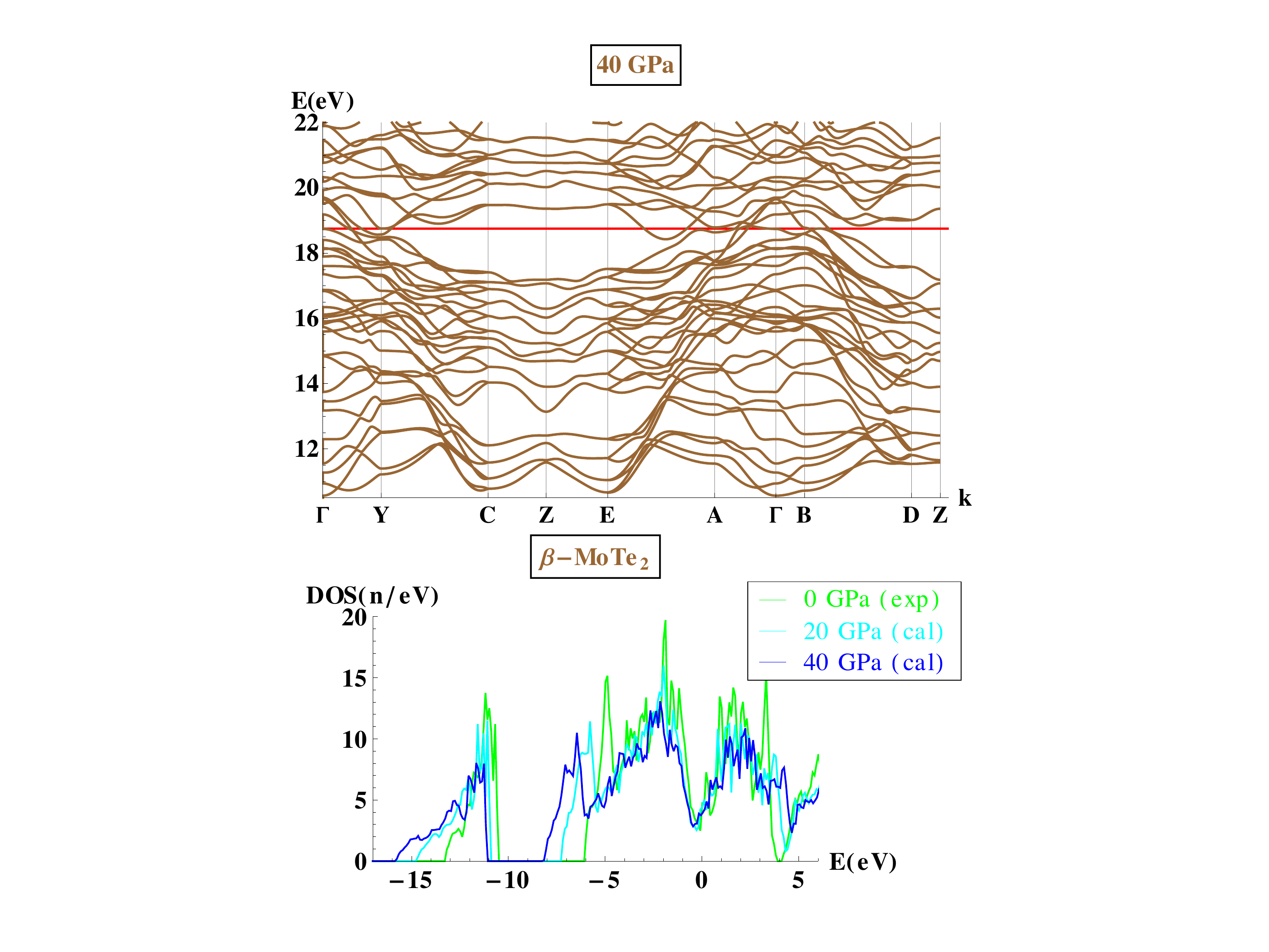}
  \caption{Band structure of $\beta$-MoTe$_2$ at $p=40$ GPa (upper
    panel) and density of states (per unit cell) at $p=20$ GPa and
    $p=40$ GPa and for the experimental unit cell at normal conditions
    (lower panel). In the DOS graph the Fermi energy is at zero.}
 \label{fig:dos_beta_MoTe2}
\end{figure}

Finally we studied the metastable $\beta$-form of MoTe$_2$
(Fig.\ref{fig:structure_beta_MoTe2}) which is already metallic at zero
pressure. The enthalpy of this form at 20 GPa is higher by 0.12
eV/(MoTe$_2$ group) than that of the 2H$_c$ form. In order to check
how the metallicity of this phase evolves with pressure we compressed
and relaxed this metastable structure to 20 and 40 GPa and
recalculated electronic DOS (Fig.\ref{fig:dos_beta_MoTe2}). The band
structure at 40 GPa is also shown in
Fig.\ref{fig:dos_beta_MoTe2}. Comparing the DOS under pressure to that
calculated for the experimental cell at normal conditions (from
Ref.\cite{Dawson_Bullett}) we see that the effect of pressure 
again does not raise much the DOS at the Fermi level.  This suggests
that pressure is not a likely tool for a major increase of
metallicity, and of BCS superconductivity, of $\beta$-MoTe$_2$.

\section{Conclusions}

The structural and electronic properties of MoSe$_2$ and MoTe$_2$ are
studied theoretically under high pressure.  Unlike MoS$_2$, these TMD
layered compounds are not prone to layer sliding transition from 2H$_c$
to the 2H$_a$ polytype.  Since MoSe$_2$ is also a good lubricant as is
MoS$_2$~\cite{Tribological_properties}, one can conclude that easy
interlayer sliding in the high pressure pristine crystals may not be
directly related to the lubricity shown by the technical emulsions
based on these materials.  Based on DFT-PBE calculations both
compounds are predicted to metallize via closing of an indirect gap
and consequent band overlap at pressures between 28 and 40 GPa
(2H$_c$-MoSe$_2$) and 13 and 19 GPa (2H$_c$-MoTe$_2$). Beyond the
metallization point they should behave as semi-metals, retaining a low
density of states at the Fermi level.  Even in the metastable
$\beta$-MoTe$_2$ phase, which is metallic already at zero pressure,
compression does not appear to increase metallicity too much. Neither
MoSe$_2$ nor MoTe$_2$ seem therefore likely, in their pristine
non-intercalated state, to become good BCS superconductors in the
range of pressures considered. Weak superlattice structural and
electronic extra Bragg spots should be looked for just below the
metallization pressure, with their possible presence providing evidence 
of an excitonic insulator state or CDW phase. In that case, enhanced 
superconductivity could arise in the metallic phase following the CDW
at higher pressures, as seen for example in 1T-TiSe$_2$~\cite{Kusmartseva}.

\begin{acknowledgments}
M.R. was supported by the Slovak Research and Development Agency under
Contract No.~APVV-0036-11. R.M. was supported by the Slovak Research
and Development Agency under Contract No.~APVV-0558-10 and by the
project implementation 26220220004 within the Research $\&$
Development Operational Programme funded by the ERDF.  Part of the
calculations was also performed in the Computing Centre of the Slovak
Academy of Sciences using the supercomputing infrastructure acquired
in project ITMS 26230120002 and 26210120002 (Slovak infrastructure for
high-performance computing) supported by the Research $\&$ Development
Operational Programme funded by the ERDF.  Work in Trieste was partly
sponsored by EU-Japan Project LEMSUPER, by Sinergia Contract
CRSII2$_1$36287/1, and by ERC Advanced Grant 320796 - MODPHYSFRICT.
\end{acknowledgments}

\end{document}